# Multiphase Interpolating Digital Power Amplifiers for TX Beamforming

Zhidong Bai, *Student Member, IEEE*, Wen Yuan, *Member, IEEE*, Ali Azam, *Student Member, IEEE*, and Jeffrey S. Walling, *Senior Member, IEEE*

*Abstract*— This paper presents a 4-channel beamforming TX implemented in 65nm CMOS. Each beamforming TX is comprised of a C-2C split-array multiphase switched-capacitor power amplifier (SAMP-SCPA). This is the first use of multiphase interpolation (MPI) for beam-steering. This technique is ideal for low-frequency beamforming and MIMO, as it does not require passive or LO based phase shifters. The SCPA is ideal to use as the core element since it can perform frequency translation, data conversion and drive an output at high power and efficiency in a compact die area. A prototype 4-element beamforming TX, occupying 2 mm×2.5 mm, can achieve peak output power of 24.4 dBm with a peak system efficiency (SE) of 24%, while achieving < 1° phase resolution and <1 dB gain error. When transmitting a 15 MHz, 64 QAM long-term evolution (LTE) signal it outputs 18.4 dBm at 14% SE with a measured adjacent channel leakage ratio (ACLR) < -30 dBc and error vector magnitude (EVM) of 3.27 %-rms at 1.75 GHz. A synthesized beam pattern based on measured results from a single die achieves <0.32°-rms beam angle error and <0.15 dB rms beam amplitude error.

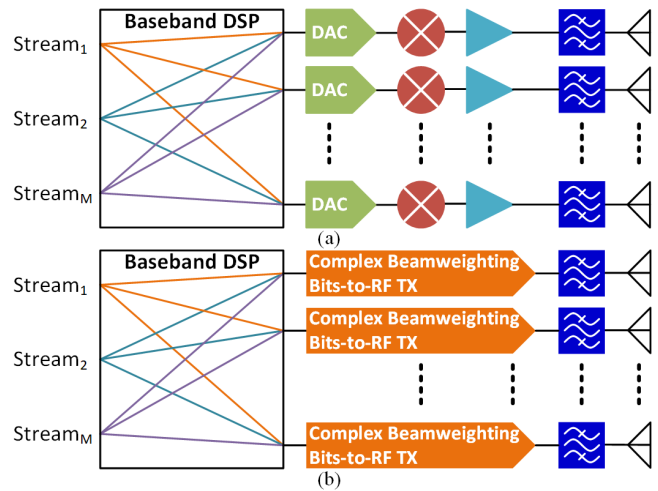

Fig. 1. Digital beamformer using (a) traditional up-converting transmitter and (b) beamweighting direct-digital bits-to-RF transmitter.

## I. INTRODUCTION

[1] In recent years, there has been tremendous growth in research focused on increasing data transmission capacity. Beamforming and MIMO techniques that leverage arrayed transceivers can increase communications capacity through SNR enhancement, spatial diversity or spatial multiplexing. The larger the array, the more that transmission capacity can be increased. In a transmitter (TX) beamformer, the amplitude and phase of the signal being transmitted on each TX in an array can be set accurately to steer a beam toward a user, or multiple beams to multiple users. TX beamformers are most flexible when individual data streams can be formed and combined in the digital baseband. Ideally, the combined spatial streams would then be connected to every antenna element in the array. However, this requires data conversion in each signal path, in addition to frequency translation and front-end amplification, as shown in Fig. 1(a). This leads to high power consumption per TX chain and precludes the use of large-scale antenna arrays.

Traditional analog beamforming can be grouped into three main methods: LO phase-shifting [1], [2], IF phase-shifting [3] and RF phase-shifting [4], [5] architectures. It is noted that TX and RX beamforming operation typically have reciprocal behavior, for each of the cases, and in many cases can share/re-use hardware. Generally, RF phase-shifting architectures consume lower power, since the baseband and IF stages can be shared for all signal paths; hence only the phase-shifting and gain weighting needs to be duplicated. However, non-linearity and losses in the can have severe impacts on the fidelity and efficiency of the TX. Additionally, purely RF and LO based approaches can only be used for single-beam beamforming. To fully leverage MIMO approaches, IF based beamforming must be used, but this requires a digital-to-analog converter (DAC), as well as frequency translation in every signal path, as shown in Fig. 1(a).

Direct-digital, bits-to-RF (DDRF) transmitters are attractive for use in digital beamforming transmitters, as they combine the functionality of a DAC and mixer [6], and can also embed the power amplifier, as in the switched-capacitor power amplifier (SCPA) [7]. Recently, DDRF techniques have begun to be investigated in digital beamforming applications [8]–[10], as

Manuscript was submitted on Sept. 14, 2019. This work was supported in part by the U.S. National Science Foundation under grant #NSF-1508701.

Z. Bai. was with the University of Utah ECE Department, Salt Lake City, UT 84112 USA. He is now with Unisoc, San Diego, CA 92121 USA (e-mail: zhidong.bai@utah.edu).

W. Yuan. is with ASML, Inc., San Jose, CA 95131 (e-mail: wen.yuan@utah.edu).

A. Azam. was with the University of Utah ECE Department, Salt Lake City, UT 84112 USA. He is now with Intel Corp., Hillsboro OR 97124 USA (e-mail: ali.azam@utah.edu).

J. S. Walling was with Microelectronic Circuits Centre Ireland (MCCI), Cork City, T12 R5CP, Ireland. He is now with Qualcomm Technologies, San Diego, CA 92121 (e-mail: jeffrey.s.walling@gmail.com).

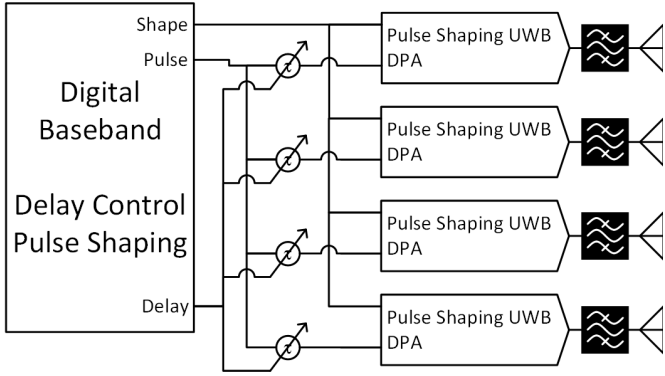
Fig. 2. DDRF used in UWB digital beamformer [8].

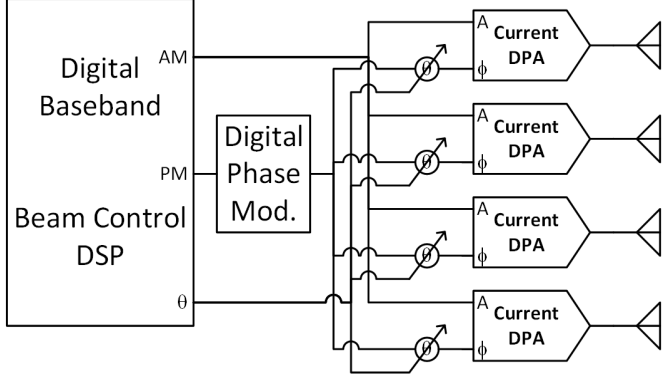
Fig. 3. DDRF used in a polar digital beamformer [9].

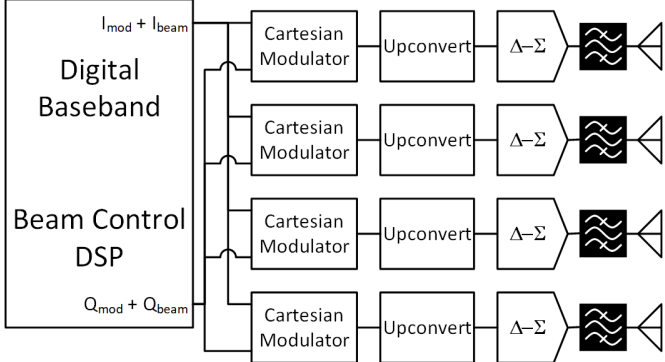
Fig. 4. DDRF used in a quadrature Δ–Σ beamformer [10].

shown in Fig. 1(b).

In this paper, a DDRF beamformer using the split-array multiphase SCPA (SAMP-SCPA) is proposed [11]–[14]. The proposed digital beamforming TX leverages the SAMP-SCPA as the core transmitter element in a 4-element beamforming transmitter, though any digital power amplifier (DPA) could be adapted to use the MP approach [15]–[20]. As with other DDRF techniques, SAMP-SCPA allows for simultaneous frequency translation, digital-to-analog conversion and front-end power amplification. Moreover, they allow for high resolution complex beamweigthing by leveraging a precision control of the clocking edges and the ratioed capacitance of a switched-capacitor array. In fact, the mechanism for beamweighting is the same mechanism used for precision wideband data modulation, with only changes to the digital baseband required for beamforming operation.

This work expands upon our conference paper, providing new details, including review of prior DDRF beamforming transmitters, analysis of the phase and amplitude accuracy and new circuit and measurement details. In section II a review of recent applications of highly digital beamforming transmitters/PAs is provided. Next, in section III the theory and operation of multiphase interpolation beamforming are provided. This is followed by section IV where details of the circuit and system design are presented. Measurement results are presented in section V and are followed by conclusions in section VI.

## II. Review of Highly Digital Beamforming Transmitters/PAs

In traditional analog beamforming transmitters, the beam-weighting is performed in either the IF/baseband, LO or RF paths, using either passive phase shifters (e.g., reflective [21]), or active phase shifters (e.g., Cartesian combiner [4], ring oscillators [22], etc.). In purely analog beamformers, only one beam can be formed per array. Digital beamforming makes it possible to operate with a single beam, or with multiple beams using digital processing, but it can be costly in hardware and power consumption, as it requires a DAC in every transmitter path.

Recently, DDRFs have been proposed for use in beamforming transmitters, as they combine the functions of data conversion, frequency translation and front-end amplification [23]. Such configurations naturally enable digital beamforming, as individual beams can be combined digitally in the baseband DSP and directly input to the DDRFs digital decoders, but to this point, there have been limited reports of DDRFs used in beamforming.

A pulse-based transmitter with digitally programmable pulse shape and delay elements was proposed by Wang, *et. al.*, for a UWB transmitter, Fig. 2 [8]. In this transmitter, digitally programmable delay lines were used to linearly adjust the true time-delay between transmitter paths. Each of the paths is comprised of a cascade of digitally programmable delay elements. This implementation does not allow complex weighting between elements, as only the delay and pulse shape can be adjusted. Though using time delay, rather than phase shifting allows for a wider frequency bandwidth, the technique suffers from reduced phase resolution at higher frequency due to the performance of the delay elements degrading.

A Cartesian phase shifter [4] was used in combination with a current-mode polar DPA by Qian, *et. al.*, to realize a DDRF beamforming transmitter, Fig. 3 [9]. In this implementation, the phase shift required for beamforming is performed using a digitally weighted Cartesian phase shifter, before driving the RF input of a current-mode DPA. Amplitude weighting could be accomplished separately in the AM path, with the caveat that higher resolution would be needed to allow for both modulation and weighting. This implementation achieves outstanding phase accuracy, at moderately high output power, owing to the polar DPA acting as an embedded transmitter, but the current-mode DPA requires digital pre-distortion (DPD) to correct for non-linearity in the DPA. Operation in the polar domain affords high energy efficiency but suffers from challenges due to systematic non-linearity.

A bandpass Δ–Σ modulator (DSM) combined with an N-path filter [10] was proposed by Zheng, *et. al.*, Fig. 4. The inherently linear 1.5b Δ–Σ DAC combined with the N-path filter reduces the transmitter area, while not requiring significant DSP. The architecture is promising, particularly for dense arrays in deeply scaled CMOS. Chip area reduction is critical for dense arrays, but it is notable that the DSM requires significant interpolation for oversampling and the digital baseband consumes a fairly high amount of power when considering that the implementation does not include a high-power output stage. The implementation achieved excellent beam accuracy, but the bandwidth (<1 MHz) and error vector magnitude (EVM) were relatively limited (>3.5-% @400 MHz, >6.6% @1.2GHz; both for single carrier 64-QAM) and would likely require significantly increased power for more complex modulation.

To overcome the challenges of the aforementioned DDRF beamforming architectures, we proposed multiphase interpolation beamforming using the SAMP-SCPA [12], Fig. 5. The SAMP-SCPA can output any required amplitude and phase, at the rate that its decoders can operate. Multiphase signaling does not have the same inherent bandwidth limitations that polar signaling does, and the SCPA is more linear than current-mode DPAs due to the use of ratioed capacitors. The addition of beamforming to the operation requires only modification to the logic decoder, which can perform the vector combination of modulation and beam-weight. Finally, the SCPA allows for an embedded output-stage/power amplifier that can operate linearly at low power [24], and at powers exceeding 1 W [16], [25]. Multiphase interpolation beamforming using an SAMP-SCPA is discussed in detail in the next sections.

III. MULTIPHASE INTERPOLATION BEAMFORMING

The proposed beamforming transmitter block diagram schematic is shown in Fig. 5. An SAMP-SCPA forms the core [14] of the multiphase interpolation beamforming transmitter. In the proposed transmitter, a multiphase (MP) clock is generated using a ring oscillator that is injection locked to a global clock. Any conventional technique can generate the MP clock (e.g., multi-stage ring oscillator, MP DLL, polyphase RC filter, etc.) The MP is then passed to an MP logic decoder that selects the appropriate phase and encodes the appropriate amplitude of each phase in the digital domain, before the signal is re-constructed using the SAMP-SCPA as a power-DAC.

The SCPA is a DPA, where the original variants were all operated in the polar domain [26], [27], [7], [28]. Digital polar operation requires a coordinate rotation digital computer (CORDIC) to convert a cartesian signal (e.g., $I+jQ$) into a polar signal (e.g., $Ae^{j\theta}$), which results in bandwidth expansion of both the amplitude ($A$) and phase ($\theta$) components, due to the nonlinear conversion. Polar systems are problematic for wide bandwidth modulation due to the bandwidth truncation and timing mis-alignment that dominate the out-of-band noise and linearity [29]. Ideally, phase modulation is implemented with a phase-locked loop (PLL), but this is unsuitable for wider band modulation [30], [31]. Quadrature modulation can be used to create wideband phase modulation [32], but if performed before the output stage, still requires precise time alignment with the amplitude signal and if performed in the output stage is subject to peak output power reduction [15], [33], [34].

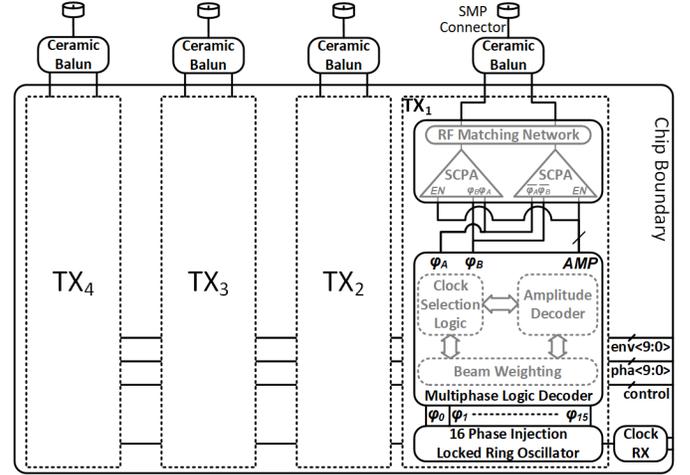

Fig. 5. Block diagram schematic of a 4-element multiphase TX beamformer using a split-array multiphase SCPA.

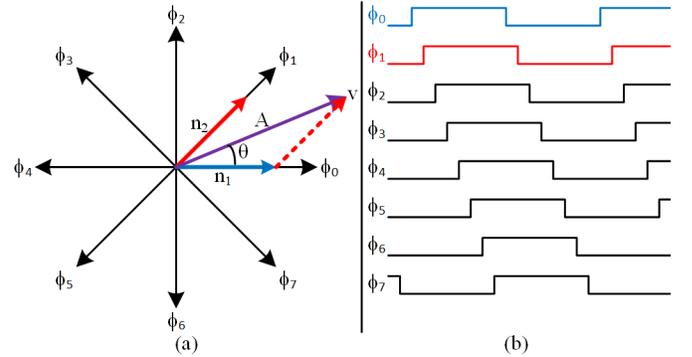

Fig. 6. Example of multiphase vector addition for 8 basis phase vectors in (a) the complex plane and (b) the associated time domain waveforms of the basis phase vectors.

The multiphase interpolator was proposed to overcome these challenges [13]. In multiphase modulation, the complex plane is subdivided into $M$-basis phases that can be weighted and summed to achieve an output at any arbitrary amplitude and phase. It should be noted that recently, the multiphase modulator has shown superior linearity as a stand-alone phase modulator [35], [36].

A. *Single Multiphase Transmitter Operation*

An example of multiphase vector addition is shown in Fig. 6, where the complex plane is divided by $M=8$ basis phases ($\phi_0$-$\phi_7$). In the example, the approach to generate the vector $v$ with amplitude, $A$, and phase, $\theta$, is depicted. First the two adjacent basis phases of the clock, $\phi_0$ and $\phi_1$, are selected, then they are individually weighted by basis phase weights, $n_1$ and $n_2$. The basis phase weights can first be found by mapping to the

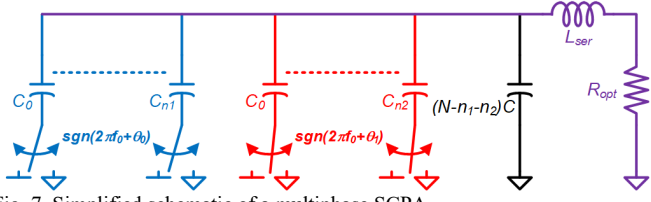

Fig. 7. Simplified schematic of a multiphase SCPA.

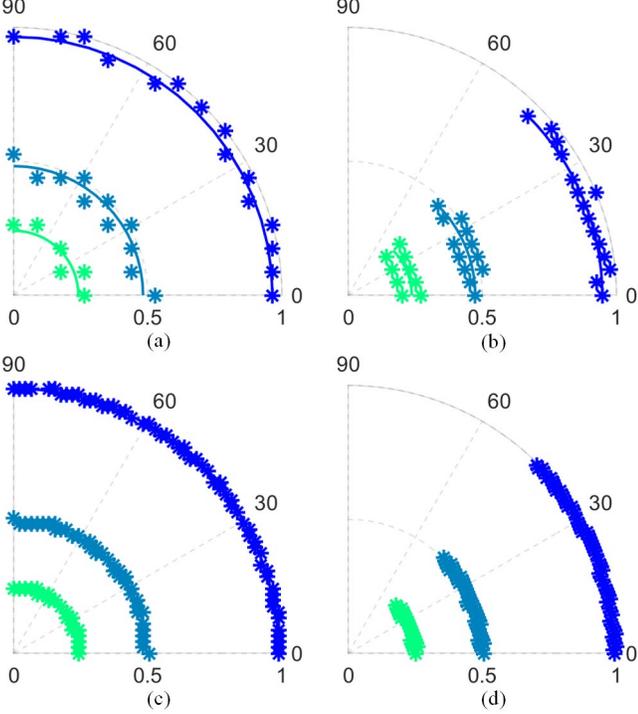

Fig. 8. Constant amplitude contours normalized to $A$=1, ¼, ½, 1/8, 0 for (a) $M$=4, $k$=3, (b) $M$=8, $k$=3, (c) $M$=4, $k$=6 and (d) $M$=4, $k$=6.

traditional Cartesian basis vectors, according to the following [13]:

$$n_1 = \frac{I\sin\left(\frac{2\pi(m+1)}{M}\right) - Q\cos\left(\frac{2\pi(m+1)}{M}\right)}{\sin\left(\frac{2\pi}{M}\right)}, \quad (1)$$

$$n_2 = \frac{-I\sin\left(\frac{2\pi m}{M}\right) + Q\cos\left(\frac{2\pi m}{M}\right)}{\sin\left(\frac{2\pi}{M}\right)}, \text{ and} \quad (2)$$

A representative multiphase SCPA (MP-SCPA) is shown in Fig. 7. The weights, $n_1$ and $n_2$, control how many of the capacitors are switched on $\phi_0$ or $\phi_1$, respectively, and the sum of $n_1$ and $n_2$ are bound by the following:

$$0 < n_1 + n_2 < 2^k, \quad (3)$$

Where k represents the total number of bits in the array (e.g., $2^k$=N). $n_1$ and $n_2$ can be found in terms of $A$ and $\theta$ using the following substation for $I$ and $Q$:

$$I = A\cos(\theta), \quad (4)$$

$$Q = A\sin(\theta). \quad (5)$$

Substitution of (4) and (5) into (1) and (2) yields the following:

$$n_1 = \frac{A \cdot \cos\left(\frac{\pi}{M}\right) \cdot \sin\left(\frac{2\pi(m-1)}{M} - \theta\right)}{\sin\left(\frac{2\pi}{M}\right)}, \quad (6)$$

$$n_2 = \frac{A \cdot \cos\left(\frac{\pi}{M}\right) \cdot \sin\left(\theta - \frac{2\pi(m-1)}{M}\right)}{\sin\left(\frac{2\pi}{M}\right)}. \quad (7)$$

It can be shown that the voltage amplitude, $V_{out}$, of the output voltage across $R_{opt}$, for a given supply voltage, $V_{DD}$, and set of codes and basis phases is given by the following [13]:

$$V_{out} = \frac{2V_{DD}}{\pi} \frac{\sqrt{n_1^2 + n_2^2 + 2n_1 n_2 \cos\left(\frac{2\pi}{M}\right)}}{N}. \quad (8)$$

The output power, $P_{out}$, is given by the following:

$$P_{out} = \frac{2}{\pi^2} \left( \frac{n_1^2 + n_2^2 + 2n_1 n_2 \cos\left(\frac{2\pi}{M}\right)}{N^2} \right) \frac{V_{DD}^2}{R_{opt}}. \quad (9)$$

The input power, $P_{in}$, is the power required to switch the total input capacitance:

$$P_{in} = C_{in} V_{DD}^2 f_0, \quad (10)$$

where $f_0$ is the output frequency. $C_{in}$ is the total input capacitance given by the following:

$$C_{in} = \left[ \frac{(n_1)(N-n_1)}{N^2} + \frac{(n_2)(N-n_2)}{N^2} + \frac{2n_1 n_2}{N^2} \right] C, \quad (11)$$

where $C$ is the value of a unit capacitor in the array. The total array capacitance can be selected to optimize matching (e.g., larger unit capacitors [37]), or efficiency at backoff by controlling the network quality factor, $Q_{NW}$ [13]. Split-array techniques allow a tradeoff between the two [14].

The SCPA is a series resonant circuit where $Q_{NW}$ is given by the following:

$$Q_{NW} = \frac{1}{2\pi NCR_{opt}}. \quad (12)$$

$L_{ser}$ (Fig. 7) is chosen to be resonant with the total array capacitance:

$$L_{ser} = \frac{1}{NC(2\pi f_0)^2}. \quad (13)$$

It is noted that the series resonant inductor and output resistor can be replaced with any load and an impedance matching network, where the aim of the transformation is to present a net inductive reactance and real impedance designed for the desired power level according to (6).

With the output power and input power known, the drain efficiency, $\eta$, can be found according to the following:

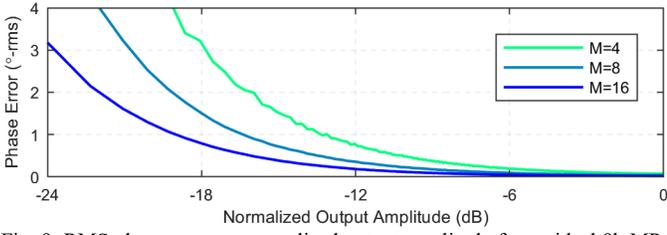
Fig. 9. RMS phase error vs normalized output amplitude for an ideal 9b MP-SCPA for varying numbers of phases, $M$.

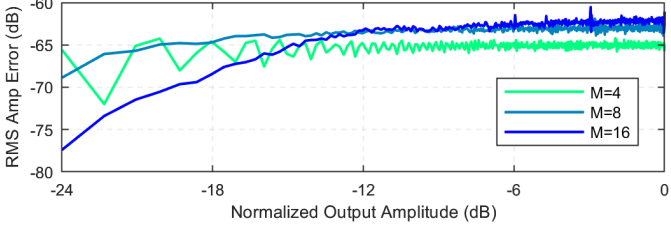
Fig. 10. RMS amplitude error vs. normalized output amplitude for an ideal 9b MP-SCPA for varying numbers of phases, $M$.

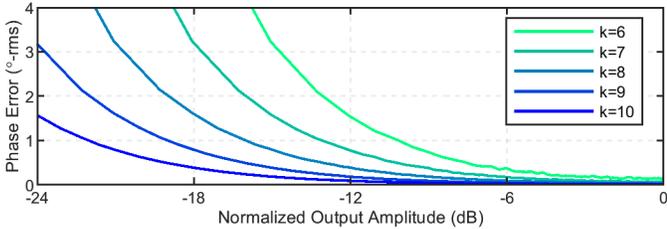
Fig. 11. RMS phase error vs. normalized output amplitude for an ideal 9b MP-SCPA for varying resolution, $k$.

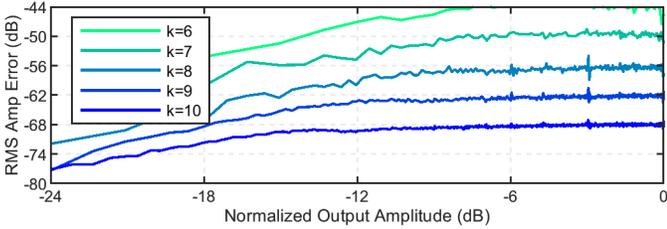
Fig. 12. RMS amplitude error vs. normalized output amplitude for an ideal 9b MP-SCPA for varying resolution, $k$.

$$\eta = \left(1 + \frac{P_{in}}{P_{out}}\right)^{-1}. \quad (14)$$

Substituting (6)-(10) into (11) yields the following:

$$\eta = \left\{1 + \frac{\pi[n_1(N-n_1)+n_2(N-n_2)]}{4 \cdot Q_{NW} \cdot \left[n_1^2 + n_2^2 + 2n_1 n_2 \cos\left(\frac{2\pi}{M}\right)\right]}\right\}^{-1}. \quad (15)$$

The efficiency can be found at any output power level and phase angle by appropriately selecting $n_1$ and $n_2$ according to (1) and (2). It is noted that choosing $Q_{NW}$ to be larger increases the efficiency level at output power backoff, at the expense of output bandwidth.

Operation using split-arrays, such as the SAMP-SCPA do not change the operation as presented above, but they do allow the resolution of the SCPA to be arbitrarily controlled while also controlling $Q_{NW}$. Additionally, it is noted that the multiphase technique originally proposed in [13] has been adapted for use only as a constant envelope phase modulator [35], [36].

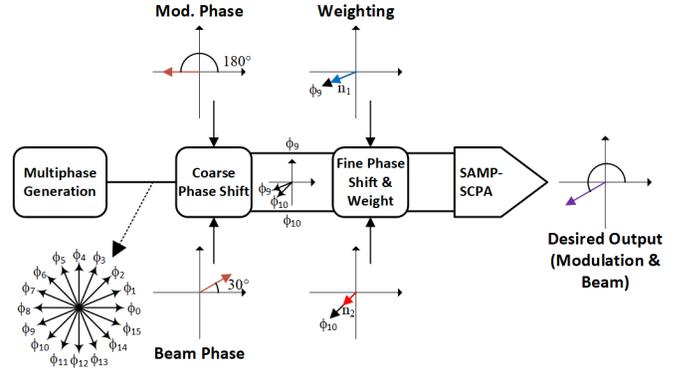
Fig. 13. Block diagram schematic of MP interpolation beamforming using an MP-SCPA.

### B. Amplitude and Phase Resolution in MP-SCPA

Both $n_1$ and $n_2$ are quantized values that will be used to reconstruct arbitrary output amplitude and phase combinations. Because number of available states decreases as the amplitude decreases, the phase resolution for small amplitudes also decreases. This is true for any digital multiphase transmitter, including the special cases of the quadrature digital transmitters (e.g., $M=4$) [15], [38]. Because the entire array can be switched fully by either $\phi_0$ or $\phi_1$, or a combination of the both, there are $2^{k+1}$ possible states between the basis phases. To illustrate, constant amplitude arcs are plotted for several values of $M$ and $k$ in Fig. 8. It is noted visually that increasing the MP-SCPA resolution, $k$, increases both the amplitude and phase resolution, as would be expected. Similarly, increasing the number of phases increases both the amplitude and phase resolution, particularly for low output amplitudes, as the density of states in each cone between two adjacent phases increases as the number of basis phases, $M$, increases.

To quantify the impact of both k and M, simulations of an ideal MP-SCPA are run across the full output amplitude and phase range. The RMS phase error is plotted as a function of the normalized output amplitude for a $k$=9b array for an ideal MP-SCPAs with $M$=4, 8 and 16 in Fig. 9. Noting that as the number of phases is increased, the discrete number of amplitude/phase states covers a reduced amount of area, meaning that the RMS error would be expected to be reduced. As expected, when doubling the number of phases, the same RMS phase error can be achieved for 3 dB less power. The RMS amplitude error is plotted as a function of the normalized output amplitude for a k=9b array for an ideal MP-SCPA with M=4, 8, and 16 in Fig. 10. Increasing the number of phases does not have significant impact on the rms amplitude error at large amplitudes, but the increased density of states does have some impact at lower amplitude.

In MP-SCPAs, it was noted that as $M$ is increased, the average power drop relative to a polar system is reduced, at the expense of reduced time available for charge settling and hence reduced linearity [13]. It was noted that $M$=16 resulted in a good tradeoff between the power drop, efficiency and linearity.

The RMS phase error is plotted versus normalized output amplitude for $M=16$ and several different array resolutions in Fig. 11. For an array resolution of 10b, < 1°-rms phase error can be achieved for ≈20 dB of output power range. The RMS amplitude error is plotted versus normalized output amplitude for $M=16$ and several different array resolutions in Fig. 12. As is expected, increasing the array resolution reduces the RMS amplitude error by ≈6dB per bit of resolution.

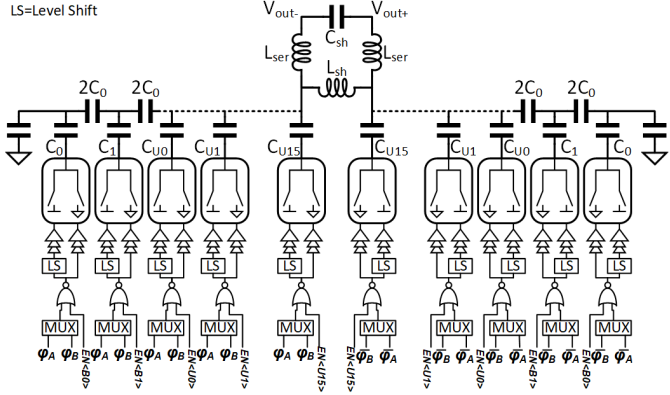

Fig. 14. Schematic diagram of differential SAMP-SCPA used in the MP TX beamformer of Fig. 2.

### C. Multiphase Beamforming Operation Example

The operation of an MP-SCPA used in MP interpolation beamforming is explained by the block diagram schematic of Fig. 13. First, a set of basis phases (e.g., $\phi_0$ - $\phi_{15}$) that span the unit circle is generated by a multiphase clock generator. Next the desired instantaneous modulation phase (180°) and beam phase (30°) are added as inputs into a phase selection logic that determines the total desired phase shift (210°) and selects two adjacent phases ($\phi_A = 202.5°$ and $\phi_B = 225°$) from the set of basis phases to this desired output phase. The desired amplitude of the output is next provided as an input as the product of the instantaneous modulation envelope and the desired beam weighting. From this, the decoder determines the $n_1$ and $n_2$, the required weighting for $\phi_A$ and $\phi_B$, respectively. Finally, these weights are applied to an SAMP-SCPA to determine how many of the cells are switched on $\phi_A$, how many are switched on $\phi_B$ and how many are held at ground. In this way, the weighted basis phases are added on the SAMP-SCPA capacitor array to form a vector summation that contains the desired output amplitude and phase modulation, as well as the desired beam-steering and weighting.

Unlike [6], which only applies digital phase shifting using a quadrature digital phase shifter at the input of a polar DPA, the proposed design allows for the amplitude and phase weighting to occur at the SAMP-SCPA and the weighting can be completely controlled by an individual multiphase logic decoder which saves power and area and allows direct recombination at the output stage.

In choosing the designed resolution, it is noted that the phase and amplitude control for the beam steering are simply added as offsets using a digital encoder. The resolution required for wideband wireless communication to suppress out-of-band noise and achieve in-band high fidelity (e.g., low EVM and ACLR) is higher than that required to achieve high phase and amplitude resolution for beam steering [14]. Hence, array resolution is primarily dictated by the in-band and out-of-band signal requirements of the communication signal to be transmitted.

## IV. CIRCUIT DESIGN DETAILS

The proposed 4-element TX beamformer architecture is shown in Fig. 5. It consists of four identical 16b TXs that each drive an off-chip ceramic balun and SMP connector jack for interface to an antenna. The resolution of the TX was chosen to be similar to the split-array MP-SCPA (SAMP-SCPA) that was previously presented in [14]. Whereas in that design, the I/O was serialized, allowing the full array resolution to be tested, in this design, the I/O remains parallelized; Hence, although the circuit is designed with a 16b array resolution, the measurement setup is limited to 22 I/O channels, meaning that only the upper 9b of the array could be utilized in the measurement setup. Circuit design details of all major blocks in the beamforming TX are now discussed.

### A. MP Clock Generation

Each TX has a local 8-stage pseudo differential ring oscillator which is injection locked to a global clock to create 16 evenly spaced basis phases. The global clock is input to the chip via an LVDS clock RX and care is taken to route the clock with equal delay to each ring oscillator to provide a common time/phase basis. In cases where layout/routing mismatch create too much phase variation, calibration can be implemented through control of the digital input codes of the individual TX slice. All phases are input into a MUX tree that is controlled by the MP logic decoder.

### B. MP Logic Decoder

The MP logic decoder takes as its input all basis phases from the MP clock generator and a digital input codes representing the desired output phase and amplitude. An additional control bit allows each transmitter to separate beam-weighting and modulation.

When the control bit is enabled, the beam weighting and steering is performed in two steps. First, the clock selection logic chooses the two adjacent phases to the desired output phases. The clock selection logic uses 4 input bits and is used to select two adjacent phases ($\phi_A$ and $\phi_B$) from the original input phases will be routed to the SAMP-SCPA. At this stage, a course phase shift has been achieved, given that the output phase must be between the two input phases that are selected. Next, using a 16b amplitude and 16b phase code the decoder finds the weights, $n_1$ and $n_2$ for $\phi_A$ and $\phi_B$, respectively, according to (1)-(7). In this way a fine beam weight and phase shift is achieved.

When the control bit is disabled the prior weight and phase shift are stored as the desired spatial weighting for the beam to be formed. The amplitude and phase input can now be added to the stored beam weighting, resulting in computation of new values for $n_1$ and $n_2$ and selection of new phases for a vector output where the output amplitude is a product of the modulation envelope with the desired beam weight and the output phase is the sum of the beam phase and modulation phase, as discussed in Section II-C.

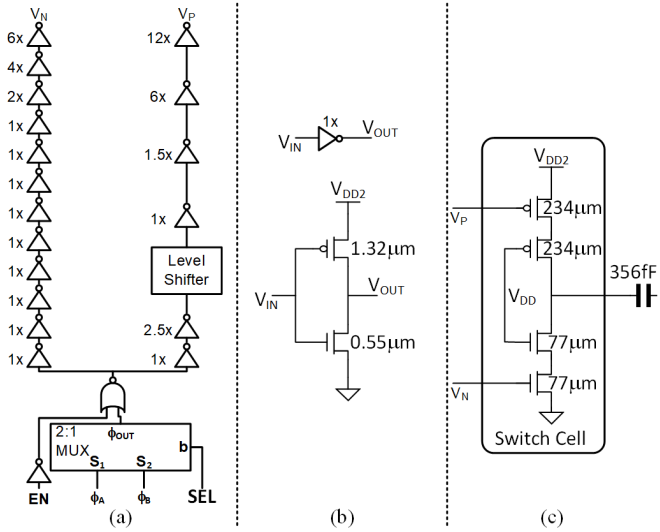

Fig. 15. (a) Block diagram schematic of the unary logic and driver slice of the SAMP-SCPA, (b) Schematic of the unit cell inverter used in the unary slice, and (c) Schematic of the output switch and capacitor used in the unary slice.

The decoders that set weights for $n_1$ and $n_2$ are identical cascaded binary-to-thermometer decoders [13]. The first 16b binary-to-thermometer decoder selects how many cells are switched by phase $\phi_A$. The second 16b binary-to-thermometer decoder selects whether the balance of cells are switched by $\phi_B$ or held at ground. All decoders are written in *Verilog* and then synthesized and automatically placed and routed.

### C. 16-b SAMP-SCPA

The schematic for one of the SAMP-SCPA cores is shown in Fig. 14. The SAMP-SCPA is chosen due to its ability to achieve good linearity and output power with high-efficiency in a compact area [14]. The schematic and layout are performed in slices, where the entire cell is designed from input logic through to capacitor as an individual slice. Unary weighted cells are designed to be identical and C-2C cells are optimized so that their delay matches the unary cells, given that each C-2C cell drives a slightly different impedance. The individual layout slices are then tiled to realize the layout of the completed SCPA. The design of each element in the slice, shown in Fig. 15, is now described starting from the input logic.

*1) SAMP-SCPA Input Logic Design*

The input to each cell of the SAMP-SCPA consists of two clock signals with identical frequency and a phase separation equal to $360°/M$, a phase selection control (SEL) that determines whether to switch on the leading or lagging phase, and an enable bit (EN) that enables/disables switching on the selected phase.

The clock signals are input into a static MUX, whose output is controlled by SEL. Any MUX implementation is appropriate, but a static NAND-NAND MUX is used for easy pitch matching with other cells in the layout. The output of the MUX drives one input of a static NOR gate, whose other input is EN. When EN is low, the clock signal propagates to the switch driver chain and when it is high, the clock signal is blocked from the driver chain. This saves power as the driver chain in each cell does not consume power when the cell is off. This also effectively closes the output switch to ground, helping the SCPA to present a constant impedance to the load.

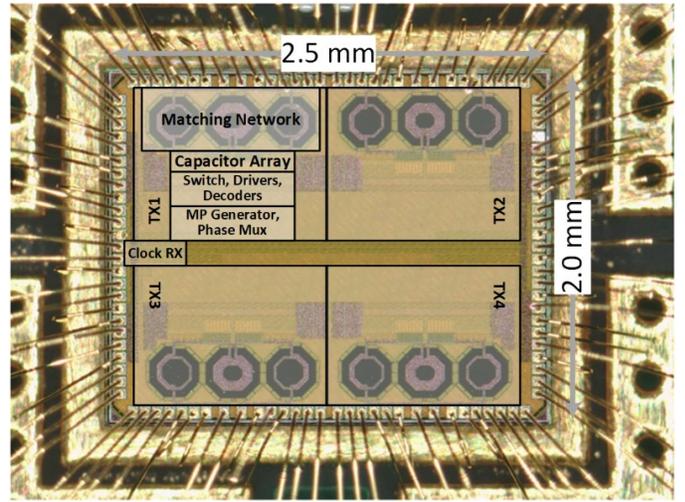

Fig. 16. Chip microphotograph of the 65nm experimental prototype 4-element beamforming TX.

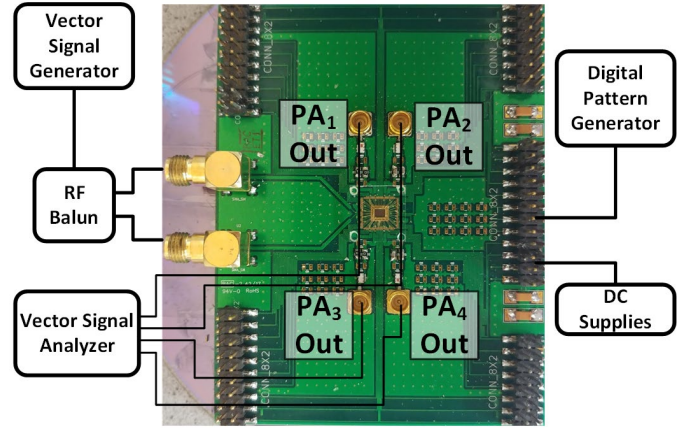

Fig. 17. Chip-on-board assembly and measurement setup.

*2) Switch Driver Slice Design*

After the input logic, a driver chain consisting of two parallel inverter chains (Fig. 15(a)) is used to drive the output switch. In one path, a level shifter like the one proposed in [39], is used to convert the input logic level from $V_{GND}$-$V_{DD}$ to $V_{DD}$-$V_{DD2}$ ($V_{DD2}=2 \times V_{DD}$). This path is used to drive the PMOS transistor in the switch (Fig. 15(c)). Inverters after the level shifters are placed in deep N-wells to allow operation from the shifted supply rails. The other path operates between $V_{GND}$ and $V_{DD}$ to drive the NMOS transistor in the switch. Each path is comprised of a cascade of scaled buffer cells based upon the unit cell, shown in Fig. 15(b).

The driver slice is located adjacent to the switch and takes its input from the MP logic decoder, where the decoders outputs have been pitch-matched to the appropriate input in the array. Co-location of the logic and driving chains allows the parasitic routing capacitance to be minimized and for easier timing synchronization of the switching signals.

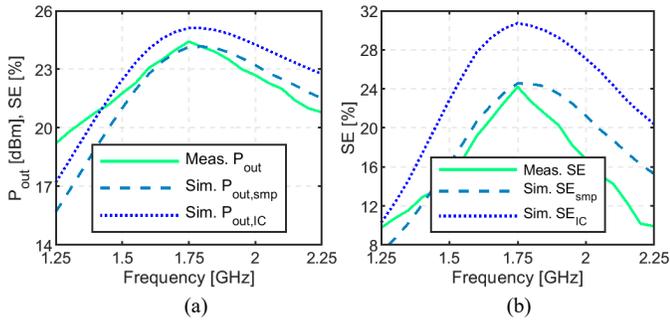

Fig. 18. (a) Measured and simulated peak output power versus frequency and (b) Measured and simulated system efficiency (SE) versus frequency.

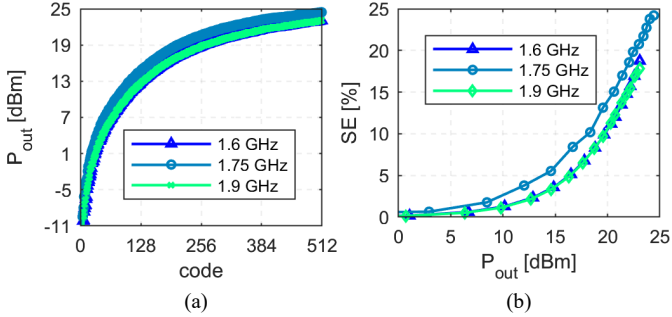

Fig. 19. Measured (a) output power versus code and (b) SE versus output power at 1.6 GHz, 1.75 GHz and 1.9 GHz.

*3) Output Switch and Capacitor*

To provide for a larger optimum termination impedance which allows for reduced loss in the output matching network, the switch is comprised of a cascoded CMOS inverter (Fig. 15(c)) This topology allows each transistor in the stack to maintain no more than $V_{DD}$ across any two terminals. This is a feature of the SCPA that is unique amongst CMOS PAs. The switch widths are optimized to drive the slice capacitance optimizing the power/delay product.

In unary MSB sub-array, all the unit capacitor in each path are identical, so the size for every switch in the unary MSB path is identical. While in C-2C LSB sub-array, the total equivalent input capacitance for each successive bit in a C-2C array increases linearly as the number of C-2C bits are increased. Also, the nodal parasitic cannot be ignored when considering the total equivalent capacitance. Hence, the size of the transistor in the switch and drivers in the C-2C LSB paths should be optimized so that the delay is matched to that of the unary paths.

The capacitors are arrayed so that the output of the switch pairs with the top plate of the capacitor, which minimizes exposure of this node to the substrate. The bottom plates of every capacitor slice in the array are shared. The array is sub-divided into a 12b C-2C LSB sub-array and 4b unary MSB sub-array. The choice of array resolution was primarily dictated by the signal fidelity requirements and complexity/area in the layout, noting that for every additional unary bit, the number of cells doubles, whereas an additional binary bit only increases the cell count by one. In our case, it has been shown that signal fidelity requirements can be largely met using 9b-10b of array resolution. The choice of where to sub-divide the array also depends on the desired linearity and complexity of the thermometer decoder required for the unary weighted bits [14].

For the presented design, the array size and segmentation considered all these options before settling on the chosen values. Using a 4b unary + 12b binary allows for a design that can meet the signal fidelity requirements and also makes possible digital pre-distortion if needed due to excess "throw-away states", all while not exceeding the assigned dimensions for each transmitter, which were dominated by I/O pad requirements and the size of the matching network, which is discussed next.

*4) Matching Network*

The total capacitance in the array seen from matching network remains constant, regardless of the input code. This is because when a switching cell is disabled, it holds the top plate of the capacitor at a constant potential through a fixed value resistance such that the impedance seen looking into each slice of the array is constant. Hence the matching network is unchanged for any choice of input code. The matching network is comprised of a shunt inductor, $L_{sh}$, a series inductor, $L_{ser}$, and a shunt capacitor, $C_{sh}$, forming a band-pass network that presents $R_{opt}$ to the array and is series resonant with the total array capacitance.

Each PA is matched to 50 Ω differentially at the pads on the chip. Bondwire inductance is in series with the PA output and is resonated with an off-chip capacitor at the center frequency of the band. An off-chip ceramic transformer balun is used to convert the output from differential to single-ended before connection to an SMP jack.

The on-chip matching and off-chip network serves to filter high frequency harmonics that arise due to the switching behavior. The -3 dB output power bandwidth is around 700 MHz, centered at 1.8 GHz. The bandwidth is determined by $Q_{NW}(\approx 3)$ of the band-pass matching network. $Q_{NW}$ is primarily chosen to maximize the efficiency of the topology while minimizing loss in the impedance transformation network. If off-chip impedance transformations are used, higher $Q_{NW}$ can be chosen.

## V. EXPERIMENTAL RESULTS

A prototype 4-element beamforming TX is fabricated in a 65 nm RF CMOS process with 9 metal layers, including an ultra-thick top metal for high quality passive components. The chip microphotograph is shown in Fig. 16. The combined area of all four TXs occupies 5mm$^2$, including the matching network, output stage, logic decoders, and the I/O pad frame; the chip area dominated by the I/O pad requirements and could be reduced in an SOC implementation.

All circuits operate from 1.4 V, except for the cascaded switches that operate from 2.8 V. The TX array is chip-on-board bonded to a PCB and an off-chip transformer balun converts the differential signal to single-ended to drive an SMP jack, as

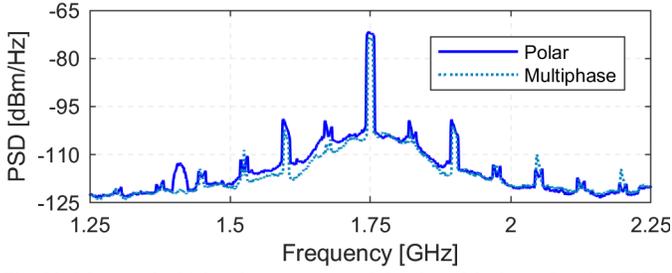
Fig. 22. Measured wideband power spectral density of the beamforming TX in both polar and multiphase modes.

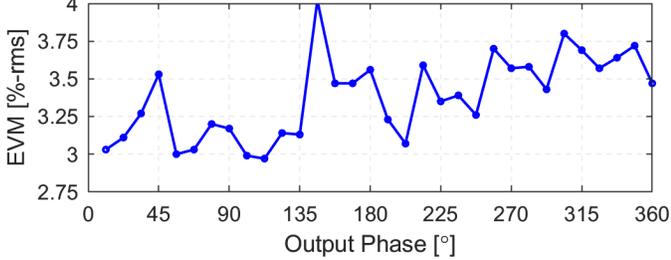
Fig. 23. Measured EVM versus output phase angle for a single transmitter.

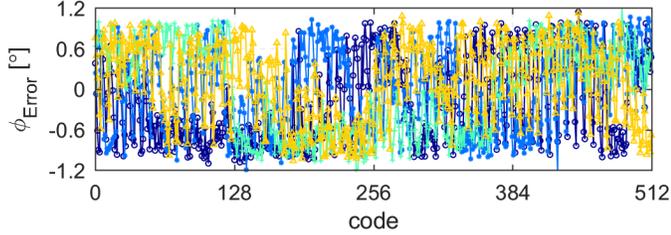
Fig. 24. Measured phase error versus output phase code for 4 elements on a single die.

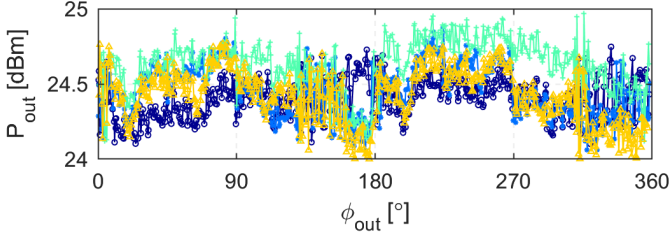
Fig. 25. Measured amplitude error vs. output phase for 4 elements on a single die.

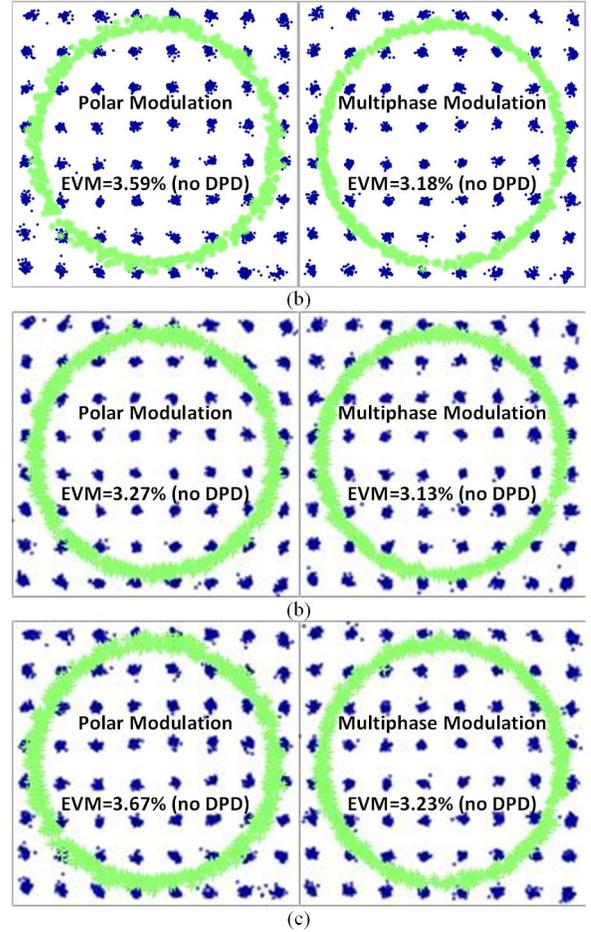
Fig. 21. Measured signal constellations for a 64 QAM, 15 MHz OFDM signal (LTE) using polar modulation (left) and for a 64 QAM, 10 MHZ OFDM signal (LTE) using multiphase modulation (right) at (a) low (1.6, 1.7 GHz), (b) center (1.75 GHz) and (c) high (1.9 GHz) frequencies.

shown in Fig. 17. An external clock signal is received by an on-chip low-voltage differential signaling (LVDS) amplifier and is used to injection lock the multistage ring oscillators in every path. The digital I/O is input from a high-speed digital I/O (HSDIO) pattern generator and is comprised of the bits to control phase selection and fine phase and amplitude control of each SAMP-SCPA. The HSDIO that was used limited the number of available I/O lines to 24, meaning that only the 9b of the array for each SAMP-SCPA could be utilized. The MSBs are favored due to their larger impact on output power and efficiency. Individual TXs are characterized for both their static and dynamic (modulated) characteristics and beamforming measurements as follows

*A. Static Measurements*

The individual TXs show similar measured performance. The measurements consider all losses including off-chip balun and SMP connector. The measured peak output power, $P_{out}$, and system efficiency (SE) versus frequency are plotted in Fig. 18(a) and (b), respectively. A peak output power and SE of 24.4 dBm and 24.2% are observed at the center frequency of 1.75 GHz, respectively. Also plotted are the simulated output power at the connector, $P_{out,smp}$, and the simulated output power at the IC periphery, $P_{out,IC}$ in Fig. 18(a). In Fig. 18(b), the simulated SE is plotted at the connector, $SE_{smp}$, and at the IC periphery $SE_{IC}$.

The measured output power versus input code and measured SE versus output power at three frequencies across the band are shown in Fig. 19(a) and (b), respectively. It is again noted that SE is measured at the connector, rather than wafer probed which is what accounts for the reduced peak efficiency when compared to other recently reported DPAs. The sharp rolloff in performance away from the center frequency is due to an external capacitor that is used to resonate the bondwire inductance from the packaging and due to a ceramic transformer balun that is placed at the output of each PA.

*B. Dynamic Measurements*

To verify the individual TX performance with signals with large PAPR, it is tested in both a polar mode where the injected clock is phase modulated and in a full multiphase mode, where the injected clock has a static input phase. In the polar modulation mode, a 15 MHz, 64 QAM, OFDM modulated signal is input to the PA, and in the multiphase mode, a 10 MHz, 64 QAM, OFDM modulated signal is input to the PA. The bandwidth was limited by the data clock buffering being undersized to drive the input parasitic at a higher desired data rate. This was due to legacy circuits that were used on the much larger 4-element chip. The problem was discovered after fabrication and limited the data clock to a rate of 150 MHz.

The measured PSD and signal constellations are shown for both the polar case and the multiphase case at low, center and high frequency in Fig. 20. The measured ACLR is <-30 dBc for all measurements. The ACLR level is largely determined by de-troughing. The signal is de-troughed to reduce the PAPR until it just meets the -30 dBc ACLR limit for E-UTRA [40]. De-troughing could be reduced to improve ACLR at the expense of reduced output power and efficiency. It is noted in the polar case that systematic non-linearities due to bandwidth limitation and timing mismatch create large spectral aliases. The measured signal constellations are shown for the polar and multiphase modulation cases at low, center and high frequencies in Fig. 21. The measured EVM for the signal is found to be < 4%-rms across all frequencies and output beam angles.

Due to the linearity of the transmitter, digital pre-distortion (DPD) is not used at any frequency and ACLR and EVM are maintained across the band. Prior work has shown that for the 64-QAM, OFDM signal used, the primary degradation to linearity is due to supply network parasitics (e.g., packaging inductance and resistance and on-chip/off-chip decoupling capacitance) [41]. To mitigate supply-network dependent non-linearity, we have adopted staggered "de-Q" decoupling capacitors to maintain a low supply network impedance across frequency, while reducing ringing due to high-Q resonances [42], [32].

The wideband power spectral density for both polar and multiphase transmission is plotted in Fig. 22. The out-of-band noise is typical for a DPA with 9b of resolution. Additionally, the signal is measured as the phase angle is arbitrarily controlled between 0° and 360°. The EVM is plotted vs output phase angle for an individual transmitter, as shown in Fig. 23.

### C. Beamforming Measurements

To validate the ability to form beams, four TXs on a single die are measured for their performance across the 9b phase control code range. The phase error is plotted as a function of phase code in Fig. 24. There is a static phase offset in the initial measurements that is corrected with a static off-line calibration. After the calibration the phase error is ≈±1° across the 9b code range.

The output power is plotted as a function of output phase in Fig. 25 for each of the four TXs and varies by <1dB for outputs across the code range. No amplitude calibration is applied.

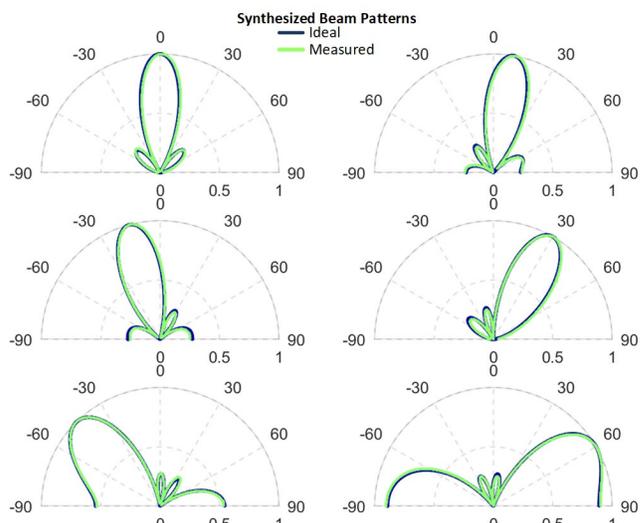

Fig. 26. Synthesized beam patterns for 4 measured TXs on a single die compared to the ideal synthesized beam pattern assuming a 4-element linear antenna array with λ/2 spacing.

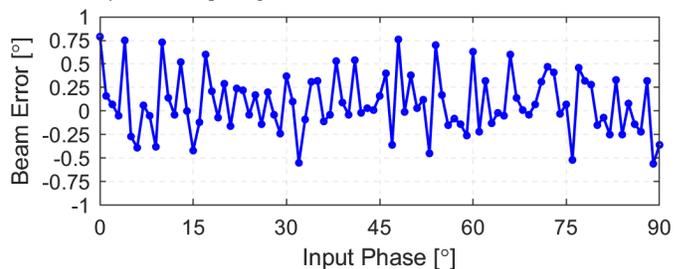

Fig. 27. Beam phase error for the synthesized array as the beam angle is steered from broadside to the horizon.

To estimate the performance of the 4 independent elements in beamforming, the array factor is synthesized using the measured data for several pointing directions for a linear array with λ/2 spacing. The same pattern is synthesized for a linear array with ideal transmitters and the two patterns are compared in Fig. 26. The synthesized beam from measured data matches well with the ideal synthesized beam across all output phases. The beam phase error as a function of ideal output phase angle is plotted for the synthesized pattern based on measured transmitters in Fig. 27. The resulting phase error across all patterns is 0.32°-rms.

## VI. CONCLUSIONS

A 4-element beamforming TX is introduced and implemented in 65nm CMOS. The beamformer leverages multiphase interpolation to enable beam-steering and beam-weighting without the need for external phase shifters or individual DACs for each element. The technique can leverage any DPA to act as the weighting element, because DPAs are bandpass transmitters that simultaneously enable frequency translation, data conversion and can operate with large effective gain in a small die area. This is because in bandpass DPAs, the output reconstruction filter is a bandpass filter that is often smaller than the low-pass baseband filters that are used in conventional up-converting transmitters. The SCPA, which in

TABLE I. COMPARISON TO RECENT DIGITAL BEAMFORMING TXS/PAS

| Ref. | Architecture | Technology (nm) | Supply (V) | Chip Area (mm$^2$) | Frequency (GHz) | Phase Resolution (°) | Phase Error (°-rms) | Pout Error (dB) |
|---|---|---|---|---|---|---|---|---|
| This Work | Multiphase Interpolation | 65 | 1.4/2.8 | 5 | 1.45-2.15* | 0.7 | 0.32 | 0.15 |
| [8] | Vernier Delay Line | 130 | 1.2 | 7.2 | 3-5 | 1 | N/A | N/A |
| [9] | Quadrature Phase Modulator | 40 | 1.1/1.2 | 8.6 | 3-7* | 0.35 | 0.2 | 0.2 |
| [10] | Quadrature Phase Modulator with ΔΣ Modulator | 40 | 1 | .19 | 0.8-1.2 | 11.25 | | |

*-3-dB P$_{out}$ bandwidth

TABLE II. COMPARISON TO RECENT DIGITAL TXS/PAS

| Ref. | Architecture | Technology (nm) | Supply (V) | Frequency (GHz) | Modulation | Linear P$_{out}$ (dBm) | Linear SE# (%) | EVM (%-rms) | ACLR (dBc) | DPD |
|---|---|---|---|---|---|---|---|---|---|---|
| This Work | SAMP-SCPA | 65 | 1.4/2.8 | 1.75 | 15 MHz, 64 QAM OFDM | 18.4* | 14* | 3.3 | -30.4/-30.8 | No |
| [14] | SAMP-SCPA | 65 | 1.2/2.4 | 1.8 | 1.4 MHz, 64 QAM OFDM | 18.9 | 21.2 | 2.65 | -30.5/-30.9 | Yes |
| [17] | Polar-IDPA | 40 | 0.5 | 2.2 | 20 MHz, 64 QAM, OFDM | 6.2 | 10.7 | 1.6 | -46/-46 | No |
| [18] | Polar-IDPA | 28 | 1.2 | 3.2 | 20 MS/s 256 QAM SC | 17.5 | 21.3$ | 2.39 | -31.3 | PM-Free |
| [19] | Polar SCPA | 65 | 3.6 | 1.9 | 5 MHz, 16 QAM, OFDM | 22.8 | 31.4$ | 5.8 | N/A | Yes |
| [20] | Polar SCPA | 40 | 1.1 | 1.5 | 20 MHz, 64 QAM, OFDM | 15.2 | 25.3 | 2.4 | -33/-30.4 | Yes |

*Includes External Losses $Drain Efficiency #SE=System Efficiency

recent years has been re-labeled an SC-RFDAC, or a C-DAC DPA, is chosen as the DPA to simultaneously realize high linearity, power and system efficiency with a small die area. When operating at 1.75 GHz, all 4 TXs can deliver a peak $P_{out}$ of 24.4 dBm with 24.2 % SE while achieving < 1° phase resolution and <1 dB gain error. The performance is validated from static and dynamic (modulation) measurements using both polar mode and multiphase mode without use of DPD. The ACLR is below the required -30 dBc LTE standard for both mode and the measured EVM is 3.27 %-rms and 3.13 %-rms, respectively.

A comparison to prior art for digital beamforming transmitters is provided in Table I. Compared to [9], which is the most closely related work, the proposed work achieves similar phase and amplitude resolution, but at higher output power and without the use of DPD, while achieving better linearity (ACLR and EVM). This partially owes to the use of the multiphase technique which does not have the systematic non-linearities of polar modulation, but primarily due to the use of the SCPA, which is more linear than the current-mode DPA. Additionally, beam amplitude weighting is natively included in the proposed work. A comparison to prior art for recent DPAs and DTXs is provided in Table II. The DPA performance alone compares well to other recent DPAs, noting that it is the only one measured to the connector, rather than wafer probed.

Because of the relatively high output power and small area, and ability to independently weight the output beam and adapt its angle on an element by element basis in the digital domain, this can be deployed in large-scale antenna arrays used for both beamforming and for MIMO.

ACKNOWLEDGMENTS

The authors wish to acknowledge the support and assistance the National Science Foundation under grant #NSF-1508701. The authors also wish to acknowledge helpful discussions with Profs. Jeyanandh Paramesh and Sangmin Yoo.